\documentclass{mn2e}
\usepackage{graphicx}
\title [RS Pup: Nebulosity and Distance]{The Nebulosity and Distance of the 
Cepheid RS Puppis}
\author[Michael Feast]{Michael W. Feast\\
Astronomy Dept, University of Cape Town, Rondebosch, 7701, South Africa\\
(email: mwf@artemisia.ast.uct.ac.za)}
\begin{document}
\maketitle
\begin{abstract}
Adopting a distance for RS Pup derived from a period-luminosity
relation based on Cepheid parallaxes, 
it is shown that the phase-lag observations of
the surrounding nebulosity by
Kervella et al. are well fitted by a model of an equatorial disc at an
angle of $ 8^{\circ}.1\pm 0^{\circ}.6$ to the plane of the sky. The
astrophysical implications of this are briefly mentioned.

\end{abstract}

\section{Introduction}
Westerlund (1961) discovered that the 41day Cepheid RS Pup was surrounded
by a remarkable nebulosity. This is in the shape of rudimentary rings but
with much distorted structure and condensations. Havlen (1972) showed that
portions of the nebulosity varied in the period of the Cepheid but with
with various phase lags. A very beautiful set of measurements of phase lags
at various points in the nebula has recently been obtained by 
Kervella et al. (2008) (= Kervella et al.). In general, the expected phase lag 
at a point $i$ may be written:
\begin{equation}
(N_{i} +\Delta \phi_{i}) =  
5.7755 10^{-3} D\theta_{i} (1+ \sin \alpha_{i})/P\cos \alpha_{i}
\end{equation}

Here $\Delta \phi_{i}$ is the fractional phase lag, $N_{i}$ the whole 
number of pulsation periods elapsed, $D$ is the distance to RS Pup in parsecs,
$\theta_{i}$ is the angular distance of $i$ from the star in arcsec,
$P$ is the pulsation period in days and $\alpha_{i}$ is the angle between the
line joining the star to $i$ and the plane of the sky (positive if $i$
is further away than the star, negative if it is nearer). The measured
quantities are $\Delta \phi_{i}$ and $\theta_{i}$. $P$ is assumed known and
here it is taken as 41.4389 days ( Kervella et al.).
In an attempt to determine $D$, Kervella et al. assume 
$\alpha_{i} = \rm {constant} =0$. That is they assume that
all the features measured by them lie in the plane of the sky
and the values of $N_{i}$ are then chosen to obtain the
best fit to this model. The
justification for this assumption is that if the nebulosity consisted of
a series of thin, uniform, spherical shells centred on the star, 
then the deviation of all measured points from the plane of the sky would
be small. However an examination of the structure of the nebulosity
(for instance from the figures in Kervella et al.) shows that it is far
from corresponding to this idealized model. There is much distortion and
density variation in the rudimentary rings. Kervella et al. place
special emphasis on the ten condensations or blobs shown in their fig 7.
The existence of such blobs is not consistent with the idealized model
and leaves open the question of whether they or other features are actually in, 
or near,
the plane of the sky. 
%Indeed it is clear from their fig. 7 that if an observer
%were viewing the system from a position in the plane of the sky roughly north
%of the star, then nine of the ten blobs marked in this figure would lie
%behind the star. Thus using their method would lead to
%an incorrect distance for the star for these blobs. 
In view of these 
uncertainties it cannot
be claimed that a definitive distance to RS Pup can be found based on 
the ``in-the-plane" assumption. In the next section this assumption is dropped
and it is shown that a simple and astrophysically interesting model for
the nebulosity is found
if a distance for RS Pup is adopted from a period-luminosity
relation.
\section{An equatorial disc model} 
van Leeuwen et al. (2007) established a reddening-free period-luminosity
relation in $V,I$ based on HST (Benedict et al. 2007) and revised Hipparcos
parallaxes. This together with the data in table A1 of that paper leads
to a predicted distance of 1728pc for RS Pup
\footnote{The distance, $1830^{+109}_{-94}$pc derived from the pulsational
parallax by Fouqu\'{e} et al. (2007) is not significantly different from
this.}.
Adopting this distance it is 
possible to
use eq. 1 to study the three dimensional structure of the nebulosity.
In principle the values of $N_{i}$ can be arbitrarily assigned. However they 
should obviously be chosen to account for apparent continuities in the
structure and to conform to some simple, physically reasonable model.
It was quickly found by trial and error that there is a consistent 
set of values of $N_{i}$
in which the points measured by Kervella et al. are further away than
the star on the south side and nearer on the north, i.e. an inclined disc
model is indicated. This is indeed the simplest model, if the
uniform spherical shell model is rejected. In such a model the values of
$N_{i}$ have to chosen such that $(N_{i} +\Delta \phi_{i})/\theta_{i}$
values are as near constant as possible in a given direction from the
star and vary smoothly with direction.
The details are given
in Table 1. This contains data on the 31 points observed by Kervella et al.
and I am greatly indebted to them for supplying detail of their observations
which were not given in the original paper. The Table lists:\\
1. Position number, $i$.\\
2. Angular distance from the star, $\theta_{i}$, in arcsec.\\
3.Azimuth of the point relative to the star, $\beta_{i}$, measured from north
through east, in degrees.\\
4. $\Delta \phi_{i}$ and its standard error.\\
5. The value of $N_{i}$ ($=N_{i}^{K}$) adopted by Kervella et al. to fit their 
model
assumptions.\\
6. The distance, $d_{y}^{K}$, behind (positive) or in front (negative) of the 
plane of the
sky through the star.
This is found by using eq.1 together with adopted values
of $N_{i}$ and $D$ to derive $\alpha_{i}$ in each case. Then,
\begin{equation}
 d_{y} = 4.848.10^{-2} \theta_{i} D \tan\alpha_{i}
\end{equation}
where $d_{y}$ is in units of $10^{-4}$pc. For $d_{y}^{K}$ 
the value of $D$ estimated by Kervella et al. (1992 pc) was combined
with their $N_{i}^{K}$ values.\\ 
7. The value of $N_{i}$ adopted in the present paper\\
8. The distance, $d_{y}$, behind or in front of the plane
of the sky through the star assumed to be at its PL distance (1728pc)
and adopting the revised values of $N_{i}$. The units are also in
$10^{-4}$pc.\\
9. The perpendicular distance $d_{x}$ of the point from the intersection of the
disc with the plane of the sky and projected onto the plane of the sky.
In the same units as $d_{y}$. 
This is given by:
\begin{equation}
d_{x} = 83.77\theta_{i} \sin(\beta_{i} -\gamma)
\end{equation}
where $\gamma$ is the angle (azimuth) at which the plane of the disc
cuts the plane of the sky.
In this test of the model this
is take as close to the line from the star to point 9
(i.e. $\gamma =  80^{\circ}$).\\  
\begin{table*}
\centering
\caption{The phase lag observations of Kervella et al. with derived linear 
positions}
\begin{tabular}{rrrrrrrrr}
\hline
$i$ & $\theta_{i}$ & $\beta_{i}$ & $\Delta \phi_{i} \pm s.e.\;\;\;\;$ & 
$N_{i}^{K}$ & $d_{y}^{K}$ & $N_{i}$ & $d_{y}$ & $d_{x}$\\
\hline
 1 & 21.10 & 139 & $0.983\pm 0.020$ & 5 &   43 & 5 & 290 & 1515\\
 2 & 21.16 & 170 & $0.809\pm 0.012$ & 5 & --24 & 5 & 233 & 1773\\ 
 3 & 21.65 & 196 & $0.989\pm 0.013$ & 5 & --7  & 5 & 252 & 1630\\
 4 & 16.58 & 186 & $0.576\pm 0.009$ & 4 & --10 & 4 & 190 & 1335\\
 5 & 16.03 & 167 & $0.504\pm 0.083$ & 4 &   19 & 4 & 210 & 1341\\
 6 & 12.89 & 213 & $0.102\pm 0.023$ & 3 & --179& 3 &--1  & 790\\ 
 7 & 10.96 & 249 & $0.098\pm 0.023$ & 3 &   19 & 3 & 148 & 175\\
 8 & 29.28 & 312 & $0.207\pm 0.036$ & 8 &   25 & 6 &--313 & --1933\\
 9 & 19.42 &  80 & $0.697\pm 0.016$ & 5 &  103 & 4 & 7   & 0 \\
10 & 17.28 & 149 & $0.602\pm 0.014$ & 4 & --70 & 4 & 146 & 1351\\
11 & 11.03 & 252 & $0.108\pm 0.004$ & 3 &   16 & 3 & 146 &  129\\
12 & 11.84 & 201 & $0.692\pm 0.010$ & 3 &  133 & 3 & 259 & 850\\
13 & 16.26 & 166 & $0.462\pm 0.001$ & 4 & --17 & 4 & 179 & 1359\\
14 & 16.45 &  67 & $0.526\pm 0.001$ & 4 & --14 & 3 & --161 & --310\\
15 & 16.98 & 148 & $0.572\pm 0.033$ & 4 & --49 & 4 & 159 & 1319\\
16 & 17.04 & 187 & $0.589\pm 0.006$ & 4 & --49 & 4 & 160 & 1365\\
17 & 17.58 &  86 & $0.394\pm 0.009$ & 4 & --178& 4 &  55 & 154\\
18 & 18.84 & 247 & $0.543\pm 0.005$ & 5 &  108 & 4 &   2 & 355\\
19 & 20.59 & 170 & $0.770\pm 0.015$ & 5 &   20 & 5 &  263& 1725\\
20 & 20.99 & 140 & $0.965\pm 0.021$ & 5 &   48 & 5 & 293 & 1523\\
21 & 24.43 & 255 & $0.925\pm 0.002$ & 6 &   50 & 5 &  15 &  178\\
22 & 24.75 &  84 & $0.182\pm 0.025$ & 6 & --252& 6 &  76 &  145\\  
23 & 26.76 & 200 & $0.460\pm 0.003$ & 7 &    12& 7 & 329 & 1941\\
24 & 28.35 & 273 & $0.081\pm 0.004$ & 7 & --289& 7 &  87 & --534\\
25 & 28.74 & 156 & $0.717\pm 0.010$ & 8 &   247& 7 & 263 & 2336\\
26 & 29.61 & 312 & $0.138\pm 0.001$ & 8 & --27 & 6 & -373 & --1955\\
27 & 30.06 &  70 & $0.158\pm 0.015$ & 8 &  --65& 7 & --28 & --437\\
28 & 33.25 & 214 & $0.570\pm 0.055$ & 9 &  117 & 8 & 190 & 2004\\
29 & 34.81 &  83 & $0.454\pm 0.006$ & 9 & --72 & 8 &  25 &  153\\
30 & 37.76 & 275 & $0.956\pm 0.004$ &10 &  163 & 8 & --48 & -819\\
31 & 39.14 & 283 & $0.938\pm 0.008$ &10 &   26 & 8 & --174 & --1281\\

\hline
\end{tabular}
\end{table*}
Fig. 1 shows a plot of $d_{x}$ against $d_{y}$. This indicates
a clear, apparently linear relation, between the two quantities. 
That is, the points lie
in a tilted plane, presumably an equatorial disc.
The line shown is a least squares
fit through the origin and is given by:
\begin{equation}
 d_{y} = 0.143(\pm 0.010)d_{x}
\end{equation}
 The tilt of the disc to the plane of the sky is $\tan^{-1} 0.143 =
8^{\circ}.1\pm0^{\circ}.6$. The rms scatter about the line in $d_{y}$ is only
$73.10^{-4}$pc,
much smaller than the diameter
of the disc, which is
$\sim 6800.10^{-4}$pc out to the limits of the phase-lag survey.
This rms scatter may be compared to the rms
scatter of $d_{y}^{K}$ which is $111.10^{-4}$pc. 
%Thus the results for
%the inclined disc at 1728pc are a better fit to the data than
%the ``in-plane" model at the Kervella et al. distance.
%It is of course
%always possible to decrease the rms scatter 
%of the in-the plane model by omitting points but
%this merely shows the defects of the original model and makes questionable
%any distance thus derived. 
No attempt has been made to optimize the disc model
by, for instance, varying $\gamma$ to find a better fit. This might reduce
the rms scatter slightly.
However this is already small compared
to the diameter of the part of the disc surveyed
indicating a relatively thin disc. 
A realistic disc will have some significant depth perpendicular to its axis
and, indeed,  Kervella et al. note that their observations at some positions
suggest smoothing attributed to
a non-zero
depth in the line of sight.
An inclined disc model broadly similar the one just discussed could
probably be derived for other distances, if there were
good evidence for these. However it should be noted that to take
the distance as a free parameter in an attempt to reduce the rms scatter
in the model is to assume that the disc must conform as closely as possible to
an idealized model which is perfectly flat and of negligible thickness.
There is no a priori justification for such an assumption.
\begin{figure}
\includegraphics[width=8.5cm]{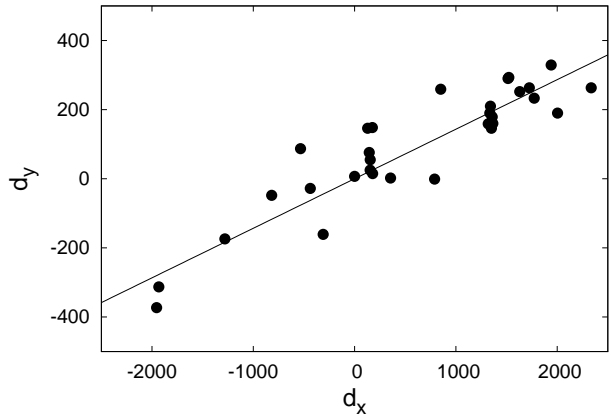}
\caption{A plot of the distances, $d_{y}$,from the plane of the sky through
the star against $d_{x}$ the perpendicular distance, in the plane of the sky,
to a line of azimuth $\gamma =80^{\circ}$. The units are $10^{-4}$pc. Note the
expanded $d_{y}$ scale.}
 
\end{figure}

An equatorial disc model for RS Pup is particularly interesting from the
astrophysical point of view. Whilst it has seemed possible, for instance,
that Cepheids might have ejected shells in a previous evolutionary phase,
it has been puzzling that only for RS Pup is such a prominent structure
found. The interpretation of the nebulosity as a disc at a small angle
to the plane of the sky opens up the possibility of a deeper understanding
of this phenomenon and its rarity. Obvious possibilities are loss
of mass in the equatorial plane by unusually rapid rotation and/or
binary interaction at an earlier evolutionary stage.
\section{Conclusion}
The structure seen in the RS Pup nebulosity makes questionable the assumption
that phase-lag observations all refer to points close to the plane of the sky
and this makes distance estimates based on this assumption questionable.
An inclined disc model at a distance
predicted by a period-luminosity relation gives a good fit to the data
and opens new possibilities for understanding the system, including a
possible interacting binary model. 
\section*{Acknowledgments}
 I am grateful to Dr Pierre Kervella for a very helpful exchange of
correspondence and to him and his colleagues for sending me detail of their
beautiful 
work not given in their paper. I am also grateful to Dr Kurt van der Heyden
for his help and to the referee for suggestions.


\begin{thebibliography}{}
\bibitem[]{} Benedict G.F. et al., 2007, AJ, 133, 1810
\bibitem[]{} Fouqu\'e P. et al., 2007, A\&A, 476, 73
\bibitem[]{} Havlen R. J., 1972, A\&A, 16, 252
\bibitem[]{}Kervella P., M\'{e}rand A., Szabados L., Fouqu\'{e} P.,
Bersier D., Pompei E., Perrin G., 2008, A\&A, 480, 167
\bibitem[]{} van Leeuwen F., Feast M.W., Whitelock P.A., Laney C.D., 2007, 
MNRAS, 379, 723
\bibitem[]{} Westerlund B.E., 1961, PASP, 73, 72
\end{thebibliography}
\end{document}